\documentstyle[sprocl]{article}

 \newcommand \be {\begin{equation}}
\newcommand	\bea {\begin{eqnarray} \nonumber }
\newcommand	\ee	{\end{equation}}
\newcommand	\eea {\end{eqnarray}}
 
 \newcommand \s	{\sigma}
\newcommand	\de	{\delta}

\newcommand	\la	{\lambda}

 \newcommand \al {\alpha}

\newcommand	\ba	{\overline}
\newcommand	\lan {\langle}
\newcommand	\ran {\rangle}
\newcommand	\Tr	{\mbox{Tr}}

\def \(	 {\left	(}
\def \)	 {\right (}

\def \form#1 {eq. (\ref{#1}) }
 
\begin{document}

\title{Statistical properties of  Random Matrices and the replica 
method}
\author{  Giorgio Parisi}
\address{Dipartimento di Fisica, Universit\`a {\em La  Sapienza} and
INFN Sezione di Roma I \\
Piazzale Aldo Moro, Rome 00185}
\maketitle

\begin{abstract}
I present here some results on  the statistical 
behaviour of large random matrices in an ensemble where the probability 
distribution is not a function of  the eigenvalues only. The perturbative 
expansion can be cast in a closed form and the limits of validity of this expansion 
are carefully analyzed. A comparison is done with a similar model with quenched 
disorder, where the solution can be found by using the replica method. Finally 
I will apply these results to a model which should describe the liquid-glass 
transition
in high dimensions.
 \end{abstract}

\section {Introduction}
I cannot to work on random matrices without thinking to to Claude Itzykson.  
I started to study this subject with him together with Eduouard Br\'ezin 
and Jean-Bernard Zuber \cite{BIPZ}.  I enjoyed very much to work in this 
group.  I had collaborated before with each of them separately, but it was 
the first time I was working with all of them together.  It was a very 
interesting and pleasant experience.  I feel still nostalgia for the long 
afternoons spent observing my friends doing long and difficult computations 
at the blackboard practically without doing mistakes.

After that work our roads partially separated.  My friends remained 
interested in the field \cite{IZ,BG}, while I started to be interested in spin glasses.  
Claude and Jean-Bernard started to study the case with two interacting matrices 
and produced after much struggle the wonderful paper on the unitary integrals 
\cite{IZ}.  At that time I had frequent discussions with Claude; he was 
explaining me the difficulties they met with their problem and I was 
telling him those I had with spin glasses.  We tried to make some 
progresses together.

This friendly exchange was quite fruitful
\footnote{I use this opportunity to 
acknowledge an other  debt  I have with Claude. I was 
 strongly influenced by his old paper with Abarbanel on 
the eikonal \cite{AI}.  I read that paper when I was starting to work in physics.  I 
learned from it the power of the functional method and the relation among 
classes of diagrams and the solution of the Schroedinger equation in a random 
Gaussian field.  That paper was quite present in my mind when ten years later we 
started to work on what we called {\sl quenched} gauge theories \cite {mpr}.}. 
Let me quote two examples.

The interpretation of the breaking of replica symmetry in terms of many pure 
states was also triggered by  Claude's observation   that
\be
\sum_b Q^s_{a,b}=
  {\prod_{i=1,s}\sum_{k_i}\lan\prod_{i=1,s}\s_{k_i}  \ran^2 \over N^s}.
\ee
A that time this relation was quite mysterious but it turned out to be a 
crucial step for later developments \cite{mpv,parisibook2}.

During one of my visits to Saclay I found a simple derivation of a result of 
Bessis \cite{BESSIS} on orthogonal polynomial.  I showed it to Claude, who was 
very interested and started immediately to derive some of the formulae needed to 
cast the result in a more compact way.  I was too busy with spin glasses so I 
didn't write it.  The results were finally incorporated in the paper on the 
unitary integrals \cite{IZ} and fortunately in this way they were  saved  from the 
oblivion.

Two years ago, when with Marinari and Ritort we started to apply the replica 
method to non random systems \cite{MPR}, it was a real pleasure for me to read 
again the paper of Claude of Jean-Bernard \cite{IZ}, which contained crucial 
results for our aims.  At that time I was thinking that one day I will come to 
Saclay to explain our results to Claude and that I will hear his comments and 
suggestions, but  sadly that day will never come.
 
The problem I will study here is the natural extension of that work of fifteen  years 
ago \cite{BIPZ} and it 
can be formulated as follows \cite{CKPR}. We have an $N 
\times N$ symmetric  matrix $M$ and we associate to it the following Hamiltonian:
\be
H(M) = N\Tr(h(N^{-1/2}M)+{1 \over N} \sum_{i, k} f(M_{i,k}).\label{HAMI}
\ee
When $N$ goes to infinity, the Hamiltonian becomes a quantity of order $N$.  We 
will consider the expectation values of intensive observables $A(M)$, which have 
a well defined limit when $N \to \infty$, e.g.  they have the same form of the 
Hamiltonian apart form a factor $N$.
 
We are interested in the computation of the equilibrium quantities
 \be 
 \lan A \ran_N
 \equiv {\int dM \exp (-\beta H(M)) A(M) \over \int dM \exp (-\beta H(M)) }
 \label{LIMIT}
\ee
in the limit where  $N \to \infty$.
 
We can also define dynamical expectation values:
we introduce the Lan\-gevin equation
\be
{dM_{i,k} \over  dt}= -{\partial H \over \partial M_{i,k}} +\eta(t)_{i,k},
 \ee
where $\eta$ is a random Gaussian distributed noise, uncorrelated for different 
pairs of indices, such that
\be
\ba{  \eta(t_1)_{i,k} \eta(t_2)_{i,k}} ={2 \over \beta} \de(t_1-t_2).
\ee
Here the bar denotes the average over different realization of the noise.

The dynamical expectation values are defined as
\be
\lan A \ran_D \equiv \lim_{t \to\infty} \lim_{N\to \infty}
\ba{A(M(t))}.
\ee
The order of the limits matters. Indeed a well known theorem states that
\be 
\lan A \ran \equiv  \lim_{N\to \infty}\lim_{t \to\infty}
\ba{A(M(t))}.
\ee
If the two limits do not commute when $A$ is the energy density, metastable 
states are present.  While metastable states with strictly infinite mean life 
cannot be present in systems with a short range interaction, the study of 
infinite range models where metastable states do exist, may be quite useful for 
understanding real systems in which there are metastable states with finite, but 
extremely large, mean life.

There are many motivations for studying this problem.

The problem is interesting {\it per se}.  It is a non-trivial generalization of 
the original problem of random matrices \cite {BIPZ} with the extra difficulty 
that the probability distribution of the matrices is no more invariant under the 
rotation group $O(N)$.  This invariance was a crucial tool in the previous 
investigations and it is not present here.

In these models at low temperature the averages in the dynamic approach may not 
coincide with those of the equilibrium approach.  Systems for which the two 
limits (infinite volume and time respectively) cannot be exchanged have a strong 
interest \cite{CUKU} and their properties may be investigated using the replica 
method \cite{FP}.
 
While there are no real doubts that in the dynamic approach the limit ($N \to 
\infty$) of intensive quantities is unique, it is possible (and in my opinion it 
is quite likely) that one can finds two (or more) different sequences $N_r$ 
(i.e. $N_r \to \infty$ when $r \to \infty$) such that the limit of the equilibrium 
quantities at low temperatures along the sequence depends on sequence.  In 
similar cases the low temperature behaviour of equilibrium properties becomes 
linked with the solution of arithmetic problems \cite{MPR,BGU}.

The techniques we will use to study the model are based on the the replica 
method.  The possibility of applying the replica method to non random systems is 
very interesting, especially because we hope that this extension may be 
useful in studying real glass.

Finally there is a model which is very interesting from the physical point: $N$ 
interacting particles constrained to move on a sphere in $\al N$ dimension
\cite{CKMP}.  This model is the simplest non trivial model for interacting 
particles which should have  a liquid-glass transition and (hopefully) can 
 be solved.
 
This paper (which is based on the results of \cite{MPR,CKPR,CKMP}) is organized 
as follows: in section 2 we present a diagrammatical analysis of the 
perturbative expansion which can be resummed in a compact way.  In section 3 we 
present a model where the Hamiltonian is random (quenched disorder).  Its 
properties coincide with those of our original model in the high temperature 
phase; the model can be solved 
analytically using the replica method.  In section 4 we discuss the relations of 
the random model with our original model and we present and disprove some 
conjectures. In section 5 we present a physically motivated model \cite {CKMP} of 
interacting particles which can be written in the form given by eq.  
(\ref{HAMI}).  Finally in section 6 we present some tentative conclusions.
 
 \section{The perturbative expansion}
 
In this section we consider the perturbative expansion for the following model 
\be 
H(M)=\frac12 \Tr(M^2) +{g_E \Tr(M^4) \over N} +g_L\sum_{1,k}M_{i,k}^4.  
\ee 
This Hamiltonian corresponds to the choice 
\bea h(x) =(\frac12 +C)x^2 +g_E x^4, 
\\ f(x) = - C x^2+ g_L x^4.  
\eea 
The value of $C$ is arbitrary, indeed 
\be 
\Tr(M^2)=\sum_{1,k}M_{i,k}^2.  
\ee 
The same term in the previous equation can be represented in both forms.  In the 
rest of the paper we will use $C=0$; in this section  will make the choice 
$\beta=1$.
 
Our aim is to investigate at the diagrammatical level if there are some 
simplifications when $N\to\infty$.  Indeed in this limit where $f=0$ (i.e.  
$g_L=0$) only planar diagrams survive.  Here the situation is more complicated.
 
A careful diagrammatical analysis shows the following \cite{CKPR} results.  

 Let us consider the contributions to a connected Green 
function and let us exclude diagrams containing self energy corrections.  In 
this case there are no mixed terms and the result is the sum of two functions, 
depending only  on $g_E$ and $g_L$ respectively.  

If we consider the self energy corrections in the internal lines, mixed terms 
may appear.  Indeed the two point function has only one possible form, 
irrespectively from the type of vertices which contribute to it (in other words, 
there is only one quadratical invariant).
 	
If we denote by $G$ the value of the propagator (I do not indicate the 
dependance on the indices), we find that any Green function ($\Gamma$) can be 
written as
\be
 	\Gamma=\Gamma_E(g_E,G)+\Gamma_L(g_L,G).
\ee

    The same conclusion is valid for the self energy diagrams, which can be 
    written as
\be
    \Sigma=\Sigma_E(g_E,G)+\Sigma_L(g_L,G).
\ee
 
     The propagator is thus given by 
\be
     G = {1 \over1+\Sigma_E(g_E,G)+\Sigma_L(g_L,G)}.
\ee
The solution of this equation gives the value of the propagator $G$ and from it 
we can reconstruct the other properties of the model.

  	These diagrammatical findings can be summarized in the following elegant way 
which generalizes to matrices the representation found  in \cite{MPR} for vectors.

 We introduce two $N  \times N$  matrices $B$ and $z$.
The total Hamiltonian for these two matrices is 
\be
H(B,z)= \Tr \left(N h(BN^{-1/2})+ \frac{R_E}2 B^2\right)+ 
\sum_{i,k}\left(f(z_{i,k})+\frac{R_L}2  z_{i,k}^2\right),
\ee
where
\be
  	    R_L=1+ \Sigma_L(g_L,G), \ \ \ \ R_E=\Sigma_E(g_L,G)
\ee
 
The previous conditions become now
\be
  		G=\lan B_{i,k}^2\ran= \lan z_{i,k}^2\ran= {1 \over R_L+R_E}.\label{FINALE}
\ee
   	
The expectation value of the energy in the original model can be finally written as
\be E= 
\lan \Tr\left (Nh(B N^{-1/2})\right)+ \sum_{i,k}f(z_{i,k})\ran 
\ee
More generally
\bea
\lan \Tr\left(N d_E(M N^{-1/2})\right)+ \sum_{i,k}d_L(M_{i,k})\ran = \\
\lan \Tr\left(N d_E(B N^{-1/2})\right)+ \sum_{i,k} d_L(z_{i,k})\ran
\eea
Under this form the result is quite compact and it can be generalized to other 
interactions which have higher powers of $M$. 

Both the integrations on $B$ and $z$ can be done explicitly: The $B$ integration 
can be done using the standard techniques, because the integrand is now 
rotational invariant. The $z$ integration factorizes in the product of $N^2$ 
independent integrals.  The values of $R_L$ and of $R_E$ can be found by solving 
the two equations (\ref{FINALE}) in two unknowns.  All computations can be done 
explicitly.

In the nutshell the final formula corresponds to an Hartree-Fock type 
approximation or, better, to two parallel spherical approximations: the local 
interaction ($f$) in the eigenvalue representation ($B$) has the only effects of 
modifying 
the quadratical term and viceversa.  This formula generalizes well known results:
\begin{itemize}
\item If $f=0$, the 
probability distribution of a given matrix element of the matrix $M$ is Gaussian.

\item If $h=0$, the probability distribution of the eigenvalues is the same of the 
Gaussian model
\end{itemize}

However we have not been able (with reasons) to obtain this result in a compact 
way.  We may speculate that the result is not valid beyond perturbation theory.  
Indeed, if such a general result would be correct, there should be a 
non-diagrammatical proof.

Generally speaking a perturbative result may break for two different reasons:
\begin{itemize}
\item
There are non analytic terms, e.g. of the form $\exp (-C/(g_E g_L))$.
\item There is a first order phase transition \footnote{In this context we say 
that 
a phase transition is of first order if no precursor signs are present, i.e. 
its presence cannot be predicted by finding  
susceptibilities which diverges when approaching the transition.}
at a non zero value of $g_E g_L$.
\end{itemize}
We shall see later that the second possibility is exactly what happens.  
However, before finding the limitations of this perturbative result, it is convenient 
to consider an other case in which the same result is obtained.
 
\section{A Random equivalent model}

It is interesting to investigate if there are other models for which the 
previous formulae are exact and can be proved in a compact way.  The model can 
be found among those with random quenched disorder.  We consider the following 
Hamiltonian:

\be H_O(z)\equiv H(z,B(z))= N\Tr\left(h(BN^{-1/2})\right)+ \sum_{i,k}f(z_{i,k}), \ee
 where $z$ is 
$N \times N$ matrix, $B(z)$ is a short notation for 
\be 
B_{i,k}=\sum_{j,l=1,N}O_{i,k;j,l} z_{j,l},\label{VINCOLO} 
\ee 
and $O$ is an $N^2 \times N^2$ orthogonal matrix, where each pair of indices 
(i.e.  $i,k$ and $j,l$) plays the role of one index.

The Hamiltonian depends on $O$ via the constraint in eq.  (\ref{VINCOLO}).  When 
the matrix $O$ is the identity, we recover the previous model.

Here we consider the case in which $O$ is a random orthogonal symmetric matrix.  
Our aim is to compute the following quantity in the large $N$ limit

\be
-N \beta F(\beta)=\int d\mu (O)\ln (\int dz \exp(-\beta H_O(z)).
\ee
At this end it is convenient to consider the quantity
\be
- n N \beta F^{(n)}=
\ln\left( (\int d\mu (O) \left(\int dz \exp(-\beta H_O(z)\right)^n\right).
\ee
It is trivial to check that 
\be
\lim_{n \to 0}F^{(n)}(\beta)=F(\beta).
\ee

The replica method consists in computing $F^{(n)}$ for integer $n$ and 
eventually continuing analytically the result to $n=0$.

The first step consist in writing
\begin{eqnarray}
\int dz \exp(-\beta H_O(z)) =\int dz dB d\la \nonumber \\
\exp\left(-\beta H(z,B)
+i \sum_{i,k}\la_{i,k}(B_{i,k}-\sum_{j,l=1,N}O_{i,k;j,l} z_{j,l})\right).
\end{eqnarray}

Now, a symmetric orthogonal matrix can be written as
\be
O=V D V^*
\ee
where $D$ is a diagonal matrix (in the generic case about half of the diagonal elements are equal to 
1, the other to -1) and $V$ is a generic orthogonal matrix.  Using the formulae of reference 
\cite{IZ} the integral over $V$ can be done (if $n$ remains finite when $N\to \infty$).  In this way 
the constraint is integrated over and we remain with two separate integrals 
\footnote{The Itzykson-Zuber formula applies in the case of integral over unitary matrices for all 
$N$.  Here we only need the leading term when $N$ goes to infinity and we can this apply their final 
expression to orthogonal matrices; in both case only planar diagrams survive.  The two results 
differ only to a crucial factor 2, which can be checked at the first non trivial order.}.

After a long computation, one finds that $F^{(n)}$ is given by the stationary 
point of $F^{(n)}(R_L,R_E)$. Here the $R$'s are $n \times n$ matrices and
\bea
\exp\left( -n N  F^{(n)}(R_L,R_E)\right)
=\int dz dB \exp\left( -\beta H_{eff}(z,B,R_L,R_E)\right),\\
\beta H_{eff}(z,B,R_L,R_E)=
\sum_{a=1,n}
 \beta H(z^a,B^a)+\\
\frac12 \sum_{a,b=1,n} \left( R_L^{a,b }\Tr(z^a z^b ) +\Tr \ln (R_L+R_E)
 + R_E^{a,b} \Tr(B^a B^b) \right) \nonumber
  \eea

In the perturbative region where at least one of the two functions $f$ and $h$ 
is small, the matrices $R$'s are diagonal and we recover the formulae of the 
previous chapter by looking to the stationary point of the free energy.

At low temperatures many new phenomena may appear. One of the most interesting 
is replica symmetry breaking. In this situation the saddle point is no more 
symmetric under the action of the replica group. The computations in the broken 
replica region are technically difficult, but they seem to be feasible and work in 
this direction is in progress \cite{PR}. 

It is interesting to recall that the the original model is one of those we 
are considering here, i.e. it corresponds to $O=1$.
  
\section{A bold conjecture}
\subsection{A naive conjecture}
The simplest conjecture would be to assume that the original formulae are 
always valid.  In order to test this conjecture, we could study the model in 
some limiting case.  It is convenient to consider a very simple non trivial 
case, i.e. matrices $M$  constrained to have matrix
elements of modulus 1;
\be
	M_{i,k}=\pm 1
\ee
This can be realized in the previous setting by using a function $h$, such that
\be
\exp(-h(z))= \delta(z^2-1)
\ee
We can now set $g_E=1$ and write the partition function of the model as
\be
\sum_{M_{i,k}=\pm 1} \exp (-\frac{\beta}{N} \Tr (M^4)).
\ee
According to the previous considerations, we have to study an auxiliary model with
partition function
\be
\int d M \exp\left( -R \Tr(M^2) -\frac{\beta}{N} \Tr (M^4)\right) \label{MODEL}
\ee
where $R$ has been chosen in such a way that $\lan B_{i,k}^2\ran =1$. In other 
words the spherical and the Ising model coincide in the high temperature region.

The solution of this problem can be found directly using the formulae of 
\cite{BIPZ} for small positive $\beta$ and their extension to the case of 
non-connected support of the eigenvalue distribution for large $\beta$.
 
Is this solution correct? 

Certainly not for negative $\beta$.  As noticed by Zinn-Justin, in this case we 
recover some form of $Z_2$ lattice gauge theory. For negative $\beta$ the 
unconfining solution (i.e.  $M_{i,k}\approx 1$) gives an energy proportional to 
$N^2$.  This is not a serious contradiction, as far also the corresponding model eq.  
(\ref{MODEL}) does not exist for negative $\beta$.

The most serious troubles appear when $\beta$ becomes large and positive.
Here we can compute the free energy of the model $F(\beta)$ and from it we can 
recover other thermodynamical quantities. It is possible to compute the entropy.

One finds (as expected in a spherical model) that  the entropy density 
$S(\beta)$ behaves at large $\beta$ as
\be
S(\beta) \approx -\ln(\beta) +const
\ee

The entropy becomes negative at sufficiently high $\beta$.  This is  
impossible; the spherical and the Ising model cannot coincide in the low 
temperature region.  Consequently, the solution of this {\sl soluble model} must 
be wrong, in the same way as in the Sherrington-Kirkpatrick model \cite{mpv}.

Something better must be devised.  Numerical simulations of the model show a 
perfect agreement for not too large values of $\beta$.  The deviations appears 
suddenly at a value of $\beta$ at which the specific heat is nearly 
discontinuous, strongly suggesting the presence of a phase transition.

 \subsection{The simplest reasonable conjecture}
 
A better conjecture, which does not go against the positivity of the entropy, 
consists in assuming that the model with $O=1$ coincides with the one with 
random $O$ also in the low temperature phase.  It is quite evident that in the 
case of random $O$ the entropy cannot be negative and this observation is enough 
to remove the entropy crisis.
 
This conjecture may be extended also to the dynamic averages, which can now be 
computed by using an appropriate formulation of the replica method \cite{FP}.
 
In this context, the analytic continuation of the free energy from the high 
temperature region has the meaning of the {\sl annealed} value of the free 
energy, i.e.  $F^{(1)}(\beta)$.
 
In principle, we can solve the random model by using the analytic formulae of the 
previous section.  This is rather hard from the technical point of view.
Detailed computations are in slow progress.
 
A trivial bound, 
\be E(\beta)\ge 1, 
\ee 
comes from the inequality 
\be N\Tr (M^4) \ge (\Tr(M^2))^2.  
\ee 
The same inequality tells us that $E=1$ only if $\Tr(M^2)$.
 
However, some information can be obtained in a simple way: the annealed 
free energy is a lower bound to the quenched free energy and that the entropy is 
non negative.  In this way one finds that 
\be E(\beta)\ge 1.06\label{BOUND} \ee
 
An approximation  which is often quite good for these systems \cite{REM}, consists in 
assuming that the entropy is given by
\be
 \max(S_a(\beta),0),
\ee
where $S_a(\beta)$ is the annealed free energy.  In the framework of this 
approximation the previous bound is exact. 

 \subsection{A counterexample}
 
 Can we test if the previous conjecture consequently is valid for all large $N$?
 Alas, yes and the answer is negative, at least for some $N$.
 
Indeed, if $N=2^k$ for some integer $k$, it is possible to find a matrix $M$ 
whose elements are all equals to $\pm 1$ and which is the square root of the 
identity.  This proves that (at least for this sequence of $N$)  the energy 
$E(\beta)$ is equal to $1$ at low temperature.  The same result may be proved 
for many other values of $N$, but it is unclear what happens for the generic $N$ 
\cite{NT}.
 
In a related model \cite{MPR,BGU} it seems that quite different results are 
obtained depending on the arithmetic properties of $N$ and that there is a way 
of taking the infinite $N$ limit (i.e.  a sequence of appropriate values of $N$ 
going to infinity) such that the model 
becomes equal to the random model.
 
The possibility of more than one thermodynamic limit for the same model 
(depending on the sequence) is quite interesting, but it is clear that there 
exist at least one way in which the conjecture we have proposed is not valid.
 
 \subsection{A more refined conjecture}
 
The physical origin of the difference in the equilibrium properties of the 
random and of the original model is related to the existence of "small" region 
in the phase space of low energy. On the contrary, it is quite likely that the 
dynamical averages do coincide with those of the random model. These 
configurations ($C_{i,k}$) have such a low energy (they correspond in some sense to a 
crystal) that they cannot be reached in the natural evolution of a system 
arriving from the high temperature region due to the presence of infinitely 
high barriers.

At low temperature the equilibrium properties of this model have not been 
carefully studied.  Here we follow the natural choice of generalizing the 
conjectures which have been done in a similar context on a simpler model which 
can be studied in a more effective way \cite{MPR,BGU}.  

We suppose that the dynamic expectations values are the same as those in the 
random model (no contradiction is present).  On the contrary the equilibrium 
expectations are definitely different from those of the random model.  We 
suppose the system has a real first order transition (with latent heat) at some 
temperature $T_C$.  At temperatures higher than $T_C$ the two models are 
equivalent.  At temperatures lower than $T_C$ the phase equivalent to the random 
model still survive as a metastable phase.

We may think to stabilize this phase by adding a new term in the 
Hamiltonian, which has zero effect on the properties of the system in the {\sl 
random} phase, but it strongly increases the energy of the {\sl crystal} phase.
An example of such a term could be
\be
   N r \theta(\sum_{i,k}((M_{i,k}-C_{i,k})^2-q))
\ee
With an appropriate choice of $C$,$r$ and $q$, it may be possible to kill the 
transition to the crystal phase without affecting the free energy in the other 
phase.  We can thus modify the model adding a {\sl small} perturbation to the 
Hamiltonian in such a way that the its free energy coincides with the free energy 
of the random model at all temperature.

\section{Hard spheres on a sphere}
In this section we  show the existence of a physically interesting model 
which can be put in the form eq. (\ref{HAMI}). In this way we prove there is 
(al least) one
non-trivial applications of the considerations presented here.

Let us consider the following Hamiltonian
\be
H= \sum_{i,k=N} V(|x_i-x_k|^2)= \sum_{i,k=N}W (x_i\cdot x_k),\label{M1}
\ee
where the $x$'s are $N$ $D$-dimensional vectors which are confined on a the 
surface of sphere \footnote{We could also put the particle in cubic box, but we 
choose a sphere for technical reasons.}
of radius $R$. The quantity $V$ is the interparticle potential as function of 
the distance squared and
\be
W(S)=V(2(R^2-S))
\ee
is the potential as function  of the scalar product.

The usual partition function of interacting particles in a $D-1$ dimensional 
flat space
is obtained by sending 
$R\to \infty$, $N\to \infty$ at fixed density $\rho=N R^{-(D-1)}$.
We may hope that for large dimensions the model will be soluble. Here we will 
study the case in which $N$, $D$ and $R$ goes to infinity together (and also 
$V$ depends on $D$.

Generally speaking we can write the  partition function under the following form
\be
\int d\mu(S) \exp\left( -\beta \sum_{i,k=N} W(S_{i,k})\right) ,
\ee
where $S$ is $N \times N$ matrix and 
\be
\mu(S) =\int \prod_i (dx_i) \prod_{i,k} \delta(S_{i,k}- x_i\cdot x_k).
\ee
We can now substitute the hard constraint $|x_i|=R$ with a term in the 
probability distribution of the $x$ proportional to $\exp(-\omega/2 
|x_i|^2$, where $\omega$ is chosen in such a way that $\lan |x_i|^2 \ran=R^2$.

The final model is slightly different form the previous one
\be
Z= \int  \prod_i (dx_i) \exp \left(-\beta H(x) -\omega/2 
\sum_{i=1,N}|x_i|^2 \right).
 \label{M2}
\ee
In sufficiently high dimensions the two models (\ref{M1} and 
\ref{M2}) should be the same, the measure on $x$ should become concentrated 
on a sphere of fixed radius which depends on $\omega$.

Now the  measure $\mu(S)$ becomes
\be
\mu(S) =\int \prod_i (dx_i) \prod_{i,k} \delta(S_{i,k}- x_i\cdot x_k)
\exp\left(-\omega/2 \sum_{i=1,N}|x_i|^2\right),
\ee
which is invariant under $O(N)$ rotations as can be seen by an explicit computation.

The model is thus reduced to the original form eq.(\ref{HAMI}), we have just to 
obtain the correct scaling for large $N$. In other terms we have to choose the 
dependance of the various term on $N$ is such a way to obtain the needed result.

It is possible to prove by an explicit computation that the {\sl right} 
scaling is obtained when $N \to \infty$ and $D \to \infty$ together at fixed 
$\alpha \equiv N/D$. The potential $V$ should also scale in an appropriate way
in this limit.

After doing the appropriate computations one arrives to a model which can be 
explicitly solved in the low density case and that behaves in a quite similar 
model to the $M_{i,k}=\pm 1$. The details of the computation can be found on the 
original paper \cite{CKMP}.

\section{Conclusions}

We can summarize these findings in the following way.
\begin{itemize}

\item The matrix model is soluble in the high temperature phase.

\item In the high temperature phase the matrix model is equivalent to a random 
model which displays a replica breaking transition to a glassy phase.

\item The random model describes the equilibrium and the dynamical properties of 
the model both in the high temperature phase and in the glassy phase.

\item For some values of $N$ there is a  transition to a crystal phase 
at low temperature.  The properties of the crystal phase cannot easily 
investigated: they depend on the arithmetic properties of $N$.

\item It may be possible to destroy the crystal phase with a {\sl small} perturbation 
which does not affect the properties in the other phase.
\end{itemize}
Some of these results are only conjectural, some have a more solid basis and 
some have been numerically verified in simpler models \cite{MPR}.  At the 
present moment we have a coherent picture of the behaviour of the model.  it 
would be extremely interesting to find out if we can collect more evidence for 
its correctness.

\end{document}